\documentclass{SciPost}
\usepackage{braket}
\usepackage{doi}
\usepackage{mathtools}
\usepackage{framed}
\usepackage{graphicx}

\usepackage[utf8]{inputenc}
\usepackage{amsmath}
\usepackage{amssymb}
\usepackage{float}
\usepackage{adjustbox}
\usepackage{color}
\usepackage{authblk}
\usepackage{soul}
\usepackage[title]{appendix}
\usepackage[normalem]{ulem}
\usepackage{algorithm}

\usepackage{algpseudocode}

\hypersetup{
    colorlinks,
    linkcolor={red!50!black},
    citecolor={blue!50!black},
    urlcolor={blue!80!black}
}

\usepackage[bitstream-charter]{mathdesign}
\DeclareUnicodeCharacter{2009}{\,}

\DeclareSymbolFont{usualmathcal}{OMS}{cmsy}{m}{n}
\DeclareSymbolFontAlphabet{\mathcal}{usualmathcal}

\definecolor{eggplant}{RGB}{126,93,181}
\definecolor{cayenne}{RGB}{148,17,0}
\definecolor{teal}{RGB}{0,145,147}
\definecolor{blueberry}{RGB}{4,51,255}

\fancypagestyle{SPstyle}{
\fancyhf{}
\lhead{\colorbox{scipostblue}{\bf \color{white} ~SciPost Physics }}
\rhead{{\href{https://doi.org/10.21468/SciPostPhys.21.4.095}{\color{scipostblue} SciPost Phys. 21, 095 (2026)}}}

\fancyfoot[C]{\textbf{\thepage}}
}

\begin{document}

\pagestyle{SPstyle}

\begin{center}{\Large \textbf{\color{scipostdeepblue}{
Compressing multivariate functions with tree tensor networks
}}}\end{center}

\begin{center}\textbf{
Joseph Tindall\textsuperscript{1$\star$},
E. Miles Stoudenmire\textsuperscript{1} and
Ryan Levy\textsuperscript{1, 2}
}\end{center}

\begin{center}
{\bf 1} Center for Computational Quantum Physics, Flatiron Institute, New York, USA \\
{\bf 2} PsiQuantum, Palo Alto, California, USA
\\[\baselineskip]
$\star$ \href{mailto:jtindall@flatironinstitute.org}{\small jtindall@flatironinstitute.org}
\end{center}

\section*{\color{scipostdeepblue}{Abstract}}
\textbf{\boldmath{%
Tensor networks are a compressed format for multi-dimensional data. One dimensional tensor networks---often referred to as tensor trains (TT) or matrix product states (MPS)---are increasingly being used as a numerical Ansatz for continuum functions by ``quantizing'' the inputs into discrete binary digits. Here we demonstrate the power of more general tree tensor networks (TTNs) for this purpose. We provide direct constructions of a number of elementary functions as generic tree tensor networks and interpolative constructions for more complicated functions via a generalization of the tensor cross interpolation algorithm. For a range of multi-dimensional functions we show how more structured tree tensor networks offer a significantly more efficient Ansatz than the commonly used tensor train. Finally, we demonstrate how the methods introduced in this work can be used to realize a TTN-based solver for multi-dimensional, non-linear Fredholm equations.
}}

\vspace{\baselineskip}

\noindent\textcolor{white!90!black}{%
\fbox{\parbox{0.975\linewidth}{%
\textcolor{white!40!black}{\begin{tabular}{lr}%
  \begin{minipage}{0.54\textwidth}%
    {\small Copyright J. Tindall \textit{et al}. \newline
    This work is licensed under the Creative Commons \newline
    \href{http://creativecommons.org/licenses/by/4.0/}{Attribution 4.0 International License}. \newline
    Published by the SciPost Foundation.}
  \end{minipage} & \begin{minipage}{0.4\textwidth}\raggedleft
    {\small Received 2025-12-10 \newline Accepted 2026-09-10 \newline Published 2026-10-15 \newline
    \href{https://doi.org/10.21468/SciPostPhys.21.4.095}{doi:10.21468/SciPostPhys.21.4.095}}%
  \end{minipage}
\end{tabular}}
}}
}

\vspace{10pt}
\noindent\rule{\textwidth}{1pt}
\tableofcontents
\noindent\rule{\textwidth}{1pt}
\vspace{10pt}

\section{Introduction}
High-dimensional data structures are ubiquitous in the modern sciences. They have an inherent exponential scaling with the number of dimensions, making any direct ``brute force" approach to their representation quite limited. Tensor networks are a compression of high-order tensors into an interconnected collection of smaller tensors \cite{Verstraete2008, Schollwoeck2011, Orus2014, Cichocki2014, Cichocki2016, Bridgeman2017, Haegeman2017, Cichocki2017, Orus2019, Vanderstraeten2019a, Cirac2021, TensorNetworkWebsite, TensorsNetWebsite, TindallGauging2023}. When the data possesses certain low-rank structure this compression can be extremely effective and turn an exponential-scaling problem into a polynomial one. The most common tensor networks take the form of one-dimensional chains of order-three tensors known as tensor trains (TT) or matrix product states (MPS). Their effectiveness has been demonstrated for a number of scientific problems ranging from one-dimensional quantum physics \cite{White1992, White1993, Vidal2003} to modeling the spread of disease \cite{Merbis2023, Dolgov2024}.

Tensor trains also offer a somewhat unconventional numerical methodology for problems in continuous space  \cite{Oseledets2010, Khoromskij2011, Dolgov2012, Khoromskij2014, Lubasch2018, GarciaRipoll2021, Richter2021, Gourianov2022, Gourianov2022-ld, Waintal2024, Alexandrov2024}. Using an encoding of the relevant continuous variables into binary strings of length $L$, mathematical functions on a grid with spacing $\mathcal{O}(\exp(-L))$ can be represented with a train of $\mathcal{O}(L)$ order-three tensors. Such an Ansatz is commonly referred to as a quantics tensor train (QTT) and has opened up a new field of tensor train-based numerical methods. While tensor trains are known to be highly effective for smooth, one-dimensional functions \cite{lindsey2024}, higher-dimensional functions can pose significant difficulties, typically requiring much larger ranks for the tensors in the train and thus larger computational resources. Other than mathematical and computational simplicity, however, there is no reason to limit these tensor-based numerical methods to trains. Tensor networks of more complex topology---which have proven fundamental in the field of two- and three-dimensional quantum simulation \cite{Tindall2023, pavesic2024, tindall2024, Weimer2017, tindall2025dynamicsdisorderedquantumsystems, doi:10.1126/sciadv.adk4321}---offer a whole new degree of freedom, allowing more structured, complex correlations to be encoded between the underlying variables (which, in this context, are the binary digits). Several works have considered Ansatzes beyond the tensor train for representing multivariate functions. Specifically: multiple tensor trains coupled via their leading tensors \cite{Dolgov2013Paper2, Ye2023}, ``functional" tensor trains where the individual tensors in the train constitute matrix-valued functions \cite{GORODETSKY201959, Soley2022} or hierarchical tucker decompositions where the binary variables are placed solely at the leaves of the tree and correlated with each other via their parents \cite{BALLANI2013639, Grasedyck2010Polynomial}. Very few methods are available in this domain, however, for working
with tensor networks of more generic topology --- with almost all software packages and algorithms currently only applicable to quantics tensor trains \cite{fernandez2024learning}. Moreover, little is understood about how the structure of the tensor network determines its effectiveness at representing a given continuous function.

In this work we rectify this lack of information and methods for working with higher-dimensional tensor networks in the context of representing continuous functions and solving numerical problems. We focus on tree tensor networks (TTNs) as the absence of loops guarantees they can be contracted with computational resources scaling polynomially in the network parameters. First, we introduce direct constructions of several elementary functions, including polynomials, on arbitrary tree tensor networks with tensor ranks bounded independent of the network. We then describe a generalization of the tensor cross interpolation algorithm \cite{Oseledets2010V2, Oseledets2011Paper2, Waintal2024, fernandez2024learning} to \textit{any} tree tensor network --- allowing the active learning of general, multi-dimensional target functions $f(\mathbf{x})$ into a TTN format. We benchmark these methods for various functions, including physically motivated ones such as the ground state density of three coupled quantum harmonic oscillators. Our benchmarks demonstrate how, for multi-dimensional functions, more structured TTNs can be a significantly more effective Ansatz than tensor trains. Finally, we introduce a new iterative tree tensor network-based solver for Fredholm integral equations --- which are ubiquitous in the physical sciences. For kernels which can be represented as a tree tensor network with fixed internal dimensions, and for equations where the fixed-point iterative approach converges, our solver has exponentially decreasing error with the size of the tree tensor network.

All of our results were produced with the open-source Julia package ITensorNumericalAnalysis.jl \cite{ITensorNumericalAnalysis} for constructing and optimizing tree tensor network representations of continuous functions, with the tree structure freely definable by the user.

\section{Preliminaries \label{sec:preliminaries}}
\begin{figure}[t!]
    \centering
    \includegraphics[width = \columnwidth]{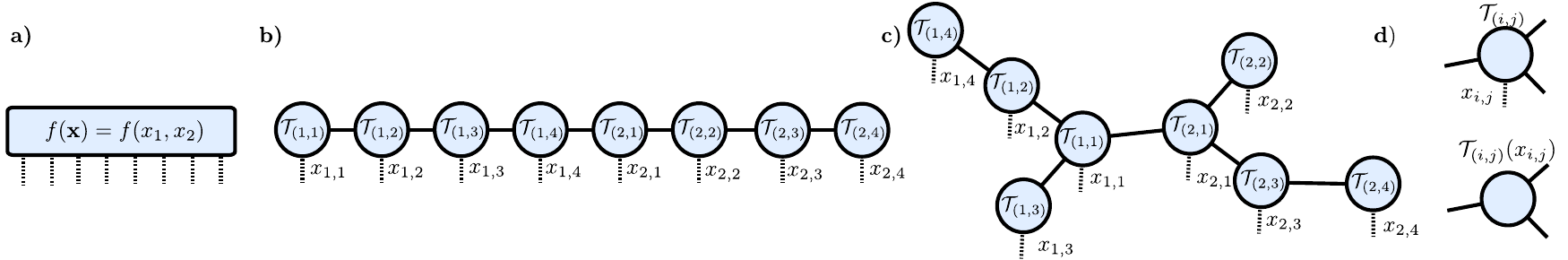}
    \caption{\textbf{a-c)} Representations of the two dimensional function $f(\mathbf{x}) = f(x_{1}, x_{2})$ with the continuous variables $x_{1}, x_{2} \in [0,1)$ encoded as binary strings $x_{1} = 0.x_{1,1}x_{1,2}x_{1,3}x_{1,4}$ and  $x_{2} = 0.x_{2,1}x_{2,2}x_{2,3}x_{2,4}$ of length four. \textbf{a)} The actual values of the function over the domain specified by the binary digits can be encoded as a single order $8$ tensor. \textbf{b}) Quantics tensor train or Matrix Product State representation of the order $8$ tensor with a binary digit ordering commonly referred to as `sequential'. \textbf{c}) An example of a tree tensor network (TTN) representation of the order $8$ tensor. \textbf{d)} Top: We use the notation $\mathcal{T}_{(i,j)}$ to refer to the local tensor in a tree corresponding to the binary digit $x_{i,j}$ -- the $j$th most significant binary digit in the decomposition of the $i$th continuous variable $x_{i}$. This tensor has $z_{(i,j)}$ `virtual' indices (black lines) connecting it to its neighbors in the tree and a single external index (dotted line) corresponding to the binary variable $x_{(i,j)} \in \{0,1\}$. Bottom: We use $\mathcal{T}_{(i,j)}(x_{i,j})$ to refer to the order $z_{(i,j)}$ tensor which is a slice of $\mathcal{T}_{(i,j)}$ for the given value of $x_{i,j}$.}
    \label{fig:ExampleTNSs}
\end{figure}

A tensor $\mathcal{T}_{v}$ is a function which maps discrete variables (which we refer to as its indices) to numbers, i.e. $\mathcal{T}_{v}(a_{1}, a_{2}, \hdots) \in \mathbb{C}$ with $a_{i} \in [1, 2, \hdots, d_{i}]$.
A tensor network is an interconnected network of tensors: each vertex of the network $v \in V$, with $V$ the set of vertices of the network, hosts a tensor $\mathcal{T}_{v}$ and the edges of the network dictate which tensors share common `virtual', `internal', or `bond' indices that are implicitly summed over.
The maximum dimension of any of the virtual (or bond) indices in the network is referred to as the bond dimension or rank $\chi$ of the tensor network. Each tensor of a tensor network can also have external indices not common to any other tensors in the network. Defining $\mathbf{I}$ as the set of all indices in the network, $\boldsymbol{\alpha}$ as the set of all indices which appear twice (i.e. are shared by two tensors) and $\mathbf{x} = \mathbf{I} \setminus \boldsymbol{\alpha}$  as the set of all external indices we have
\begin{equation}
    \mathcal{T}(\mathbf{x}) = \sum_{\boldsymbol{\alpha}} \prod_{v \in V}\mathcal{T}_{v}(\mathbf{I_{v}}), \qquad \mathbf{I_{v}}, \boldsymbol{\alpha}, \mathbf{x} \subseteq \mathbf{I},
\end{equation}
with $\mathbf{I_{v}}$ the indices possessed by the tensor $\mathcal{T}_{v}.$ The tensor network $\mathcal{T}$ thus corresponds to a factorization of a single tensor with indices $\mathbf{x}$. Some example tensor networks are illustrated in Fig. \ref{fig:ExampleTNSs}b) and c) along with the tensor they are representing in Fig. \ref{fig:ExampleTNSs}a).

In this work the external indices of the network represent discrete variables which decompose a series of $n$ continuous variables $\mathbf{x} = (x_{1}, x_{2}, ..., x_{n}) \in [0, 1)^{n}$ in a binary manner. For a given continuous variable $x_{i}$ the binary decomposition reads $x_{i} = \sum_{j=1}^{L}\frac{x_{i,j}}{2^{j}}$
where the $x_{i,j} \in \{0, 1 \}$ are the binary \textit{variables} or \textit{bits} which are each represented by an external index of dimension $2$ in the tensor network \footnote{In this work, for simplicity, we assume binary strings which are all of length $L$ and thus the total number of external indices is $nL$. The generalization of our work to $n-$ary decompositions with a variable number of binary digits for each variable is straightforward.}.
The $2^{nL}$ possible binary strings, or configurations of the bits, realize a uniform discrete grid for $\mathbf{x} \in [0, 1 - \delta]^{n}$ where $\delta = 2^{-L}$ is the grid spacing in each dimension which is exponentially `fine' in the number of bits.
Unless specified otherwise we will focus on the case where each tensor in the network contains one external index and therefore corresponds to one binary digit $x_{i,j}$ in the decomposition of a continuous variable $x_{i}$. We will focus exclusively on tensor networks which are trees, i.e. networks where the virtual indices do not form loops. We should emphasize that this is related, but not the same as the previously introduced hierarchical tucker format \cite{BALLANI2013639, Grasedyck2010Polynomial, Grasedyck2010} where the binary variables live only on the leaves of the tensor network and tensors in the bulk are simply factors that mediate correlations between them and do not contain the external variables. The fact that the tensor networks we consider do not contain loops means that they can be optimized and contracted (for a given configuration of their binary variables) efficiently and we will refer to them as tree tensor networks (TTNs).

The TTNs in this paper have a structure specified by a labeled tree $\mathcal{T}$ where each of the vertices corresponds to a single binary digit $x_{i,j}$ in the decomposition of $x_{i}$. We will use the notation $\mathcal{T}_{(i,j)}$ to denote the tensor on a given vertex and $\mathcal{T}_{(i,j)}(x_{i,j})$ to refer to a given `slice' of that tensor $\mathcal{T}_{(i,j)}$ for a specific value of $x_{i,j} \in \{0, 1\}$, effectively viewing it as a `function' of the binary variable $x_{i,j}$. 
Each $\mathcal{T}_{(i,j)}(x_{i,j})$ is just another tensor and has order $z_{i,j}$: the co-ordination number of the given vertex $x_{i,j}$ in the tensor network. This idea is illustrated in Fig.~\ref{fig:ExampleTNSs}d. We refer to the $z_{i,j}$ indices connecting a local tensor to its neighbors as `internal' or `virtual' indices. We frequently use the notation $\boldsymbol{\alpha}_{i,j} = (\alpha_{1}, \alpha_{2}, \hdots \alpha_{z_{i,j}})$ to denote the virtual indices on the tensor $\mathcal{T}_{(i,j)}$ and index them starting from $0$, i.e. $\alpha_{k} = 0,1,2, \hdots, {\rm dim}(\alpha_{k})$.

The tensor network effectively encodes the values for some function $f(\mathbf{x})$ over the uniform discrete grid $\mathbf{x} \in [0, 1 - \delta]^{n}$ with $2^{nL}$ grid points. A fixed `configuration` of its external indices uniquely specifies a value for $\mathbf{x}$. The contraction of the resulting network 
yields the scalar $\mathcal{T}(\mathbf{x})$ that approximates $f(\mathbf{x})$. 
Such a contraction can be done in $\mathcal{O}(nL \chi^{z})$ time, where $z$ is the maximum co-ordination number of any of the tensors in the network: i.e. $z = {\rm sup} \{z_{1,1}, z_{1,2}, \hdots , z_{n, L} \}$.
We illustrate two example tree tensor networks in Fig.~\ref{fig:ExampleTNSs} as decompositions of a two-dimensional function. In this work we will provide methods for constructing functions as a tensor network with \textit{any} choice of labeled tree $\mathcal{T}$, allowing us to compare the effectiveness of different tree structures. 

\section{Constructing functions as tree tensor networks}
\label{sec:Construction}
Here we detail how to construct various functions as tree tensor networks of generic topology. We first describe a direct methodology for certain functions via explicit setting of the tensor elements in the network and provide rules for adding and multiplying functions which greatly expands the space of functions which can be \textit{exactly} represented. We then provide an indirect methodology for more generic functions via the tensor cross interpolation algorithm, which variationally minimizes the infinity norm between the tree tensor network and some desired function.

\subsection*{Direct construction}

Certain elementary functions are factorizable as a product of separate functions for each bit $f(\mathbf{x}) = \prod_{i=1}^{n}\prod_{j=1}^{L}f_{(i,j)}(x_{i,j})$. 
The corresponding tensor network thus has rank or bond dimension one, with local tensors
$\mathcal{T}_{(i,j)}(x_{i,j}) = f_{(i,j)}(x_{i,j})$ with either no virtual indices or virtual indices of dimension one, allowing them to be ``trivially" represented on any desired topology. Three such classes of rank-one functions are:

\begin{itemize}
    \item \textit{Constant functions}: $f(\mathbf{x}) = c = \prod_{i=1}^{n}\prod_{j=1}^{L}c^{\frac{1}{nL}}$.
    \item \textit{Exponential functions}: $f(\mathbf{x}) = c e^{\mathbf{k} \cdot \mathbf{x} + a} = \prod_{i = 1}^{n}\prod_{j=1}^{L}c^{\frac{1}{nL}}e^{k_{i}\frac{x_{i,j}}{2^{j}} + \frac{a}{nL}}$ with $a, c \in \mathbb{C}$ \newline and $\mathbf{k} = (k_{1}, k_{2}, ..., k_{n}) \in \mathbb{C}^{n}$.
    \item \textit{Dirac delta function}: $f(\mathbf{x}) = \delta (\mathbf{x} -\tilde{\mathbf{x}}) = \prod_{i = 1}^{n}\prod_{j=1}^{L} 2\delta_{x_{i,j},\tilde{x}_{i,j}}$ where $\delta_{n,m}$ is the Kronecker delta function and $\tilde{x}_{i,j}$ is the setting of digit $(i,j)$ in the binary decomposition of $\tilde{\mathbf{x}}$.
\end{itemize}

\textit{Polynomials} - A more non-trivial case is that of polynomials of degree $d$. Here we will provide a construction of the one-dimensional degree $d$ polynomial $p(x) = \sum_{k = 0}^{d}c_{k}x^{k}$ with $c_{0}, c_{1}, \hdots c_{d} \in \mathbb{C}$ on any tree tensor network where the bond dimension will be $\chi = d + 1$, independent of the choice of tree. As we are working in one dimension we will drop the dimension subscript $i$ in our notation for $x_{i,j}$ and the local tensor $\mathcal{T}_{(i,j)}$, i.e. $x_{i,j} \rightarrow x_{j}$ and $\mathcal{T}_{(i,j)} \rightarrow \mathcal{T}_{j}$.

First we pick \textit{any} of the binary digits for the continuous variable $x$ and designate it as $x_{r}$. For $j \neq r$ the local tensor is $\mathcal{T}_{j}$ and we will denote its elements as $\mathcal{T}_{j}(x_{j})_{\alpha_{1}, \alpha_{2}, \hdots, \alpha_{z_{j}-1}, \beta}$ where $\beta$ is the virtual index corresponding to the edge which separates $x_{j}$ from  $x_{r}$ and the $\alpha_{1}, \alpha_{2}, \hdots, \alpha_{z_{j}-1}$ denote the remaining virtual indices of the tensor. For $j = r$ we define the on-site tensor $\tilde{\mathcal{T}}_{r}$ and its elements as $\tilde{\mathcal{T}}_{r}(x_{r})_{\alpha_{1}, \alpha_{2}, \hdots, \alpha_{z_{r}}}$.

The elements of the tensors in the network are then
\begin{align}
        &\mathcal{T}_{j}(x_{j})_{\beta} = \left(\frac{x_{j}}{2^{j}}\right)^{\beta} \qquad j \neq r, \ z_{j} = 1 \notag \\
        &\mathcal{T}_{j}(x_{j})_{\alpha_{1}, \alpha_{2}, \hdots, \alpha_{z_{j}-1}, \beta} = C_{\alpha_{1}, \alpha_{2}, \hdots, \alpha_{z_{j}-1}, \beta} \left(\frac{x_{j}}{2^{j}}\right)^{f_{\alpha_{1}, \alpha_{2}, \hdots, \alpha_{z_{j}-1}, \beta}} \qquad j \neq r, \ z_{j} > 1 \notag \\
        &\tilde{\mathcal{T}}_{r}(x_{r})_{\alpha_{1}, \alpha_{2}, \hdots, \alpha_{z_{r}}} = \sum_{\beta = 0}^{d}c_{\beta}C_{\alpha_{1}, \alpha_{2}, \hdots, \alpha_{z_{r}}, \beta} \left(\frac{x_{r}}{2^{r}}\right)^{f_{\alpha_{1}, \alpha_{2}, \hdots, \alpha_{z_{r}}, \beta}} \qquad
        \label{Eq:PTensors}
\end{align}
where we have introduced
\begin{equation}
    f_{a_{1}, a_{2}, \hdots, a_{n}, b} =
         b - \sum_{i = 1}^{n}a_{i}
    \label{Eq:Exponent}
\end{equation}
and
\begin{equation}
    C_{a_{1}, a_{2}, \hdots, a_{n}, b}  = \begin{cases}
         \binom{b}{a_{1}} & n = 1 \ {\rm and} \ f_{a_{1}, a_{2}, \hdots, a_{n}, b} \geq 0 \\
         \binom{b}{a_{1}}\prod_{i=1}^{n-1}\binom{f_{a_{1}, a_{2}, \hdots, a_{i}, b}}{a_{i+1}} &  n > 1 \ {\rm and} \ f_{a_{1}, a_{2}, \hdots, a_{n}, b} \geq 0 \\
         0 & {\rm Otherwise} 
         \label{Eq:Coefficient}
    \end{cases}
\end{equation}
for arbitrary integers $a_{1}, a_{2}, \hdots a_{n}$ and $b$. The $c_{\beta}$ are the coefficients of the polynomial. In the supplementary material we prove that this tree tensor network will contract to the one-dimensional polynomial $f(x)$ for any configuration of its external indices. We also describe how to elevate the construction to the multidimensional case $f(\mathbf{x}) = p(x)$ with \newline $x \in \{x_{1}, x_{2}, \hdots, x_{n}\}$ when there are external indices which decompose continuous variables other than $x$. We emphasize that our construction here is completely general and works on any tree: in the case the tree forms a one-dimensional path our result reduces to the known quantics tensor train construction \cite{Lubasch2018, lindsey2024}

\subsubsection*{Multiplication and addition}
The direct constructions above can be combined with rules for multiplying and adding together tensor networks to vastly expand the space of possible functions which can be realized. We detail these below for a generically structured tensor network, emphasizing that in the case the tree forms a one-dimensional path our rules reduce to the well-established formalism for adding and multiplying QTTs \cite{Shinaoka2023, Ripoll2024}. 

\textit{Addition} - Consider two tensor networks which are defined over the same labeled tree $\mathcal{T}$ and encode two functions $t_{1}(\mathbf{x})$ and $t_{2}(\mathbf{x})$ and have bond dimensions $\chi^{(1)}$ and $\chi^{(2)}$. We define their local tensors as $\mathcal{T}^{(1)}_{(i,j)}$ and  $\mathcal{T}^{(2)}_{(i,j)}$. The external index on a given vertex is common between the two networks (i.e. it encodes the same binary digit) but virtual indices are not. We define the `addition' of the two tensor networks as a new tensor network over the same labeled tree $\mathcal{T}$ with on-site tensors $\tilde{\mathcal{T}}_{(i,j)}$ which are defined via $\tilde{\mathcal{T}}_{(i,j)}(x_{i,j}) = \mathcal{T}^{(1)}_{(i,j)}(x_{i,j}) \oplus \mathcal{T}^{(2)}_{(i,j)}(x_{i,j})$ where $\oplus$ represents the tensor direct sum. By the definition of the direct sum, the dimension of the virtual indices of $\tilde{\mathcal{T}}_{(i,j)}$ is the sum of the two corresponding virtual indices in $\mathcal{T}^{(1)}_{(i,j)}$ and $\mathcal{T}^{(2)}_{(i,j)}$. It follows that, for a given configuration of the external indices of the new network, the contraction will yield $\tilde{t}(\mathbf{x}) = t_{1}(\mathbf{x}) + t_{2}(\mathbf{x})$ and the bond dimension of the new network is $\chi = \chi^{(1)} + \chi^{(2)}$.

\textit{Multiplication} -
Consider again two tensor networks which are defined over the same labeled tree $\mathcal{T}$ and represent two functions $t_{1}(\mathbf{x})$ and $t_{2}(\mathbf{x})$ and have bond dimensions $\chi^{(1)}$ and $\chi^{(2)}$. We define the `multiplication' of the two networks as a new tensor network over the same labeled tree $\mathcal{T}$ with resulting on-site tensors $\tilde{\mathcal{T}}_{(i,j)}$ which are defined via  \newline $\tilde{\mathcal{T}}_{(i,j)}(x_{i,j}) = \mathcal{T}^{(1)}_{(i,j)}(x_{i,j}) \otimes \mathcal{T}^{(2)}_{(i,j)}(x_{i,j})$, where $\otimes$ denotes the tensor outer product. The tensor $\tilde{\mathcal{T}}_{(i,j)}$ thus has $2z_{i,j}$ virtual indices, and one external index corresponding to the digit $x_{i,j}$. There are two virtual indices for each edge in $\mathcal{T}$ and these pairs of indices can be combined together into a single index to recover a tensor network over $\mathcal{T}$ but with the dimension of the virtual index on a given edge being the product of the dimension of the corresponding indices for that in the original tensor networks. The bond dimension of the new network is thus $\chi = \chi^{(1)}\chi^{(2)}$.
 It follows that, for a given configuration $\mathbf{x}$ of the external indices the new tensor network contracts to $\tilde{t}(\mathbf{x}) = t_{1}(\mathbf{x})t_{2}(\mathbf{x})$.

 \textit{Example} - The hyperbolic function $f(\mathbf{x}) = c \cosh(\mathbf{k} \cdot \mathbf{x} + a)$ with $a, c \in \mathbb{C}$ and \newline $\mathbf{k} = (k_{1}, k_{2}, ..., k_{n}) \in \mathbb{C}^{n}$ can be built as a tensor network over \textit{any} labeled tree $\mathcal{T}$ with $\chi = 2$ by combining the exponential definition and the rule for addition. The local tensor elements are
 \begin{equation}
     \mathcal{T}_{(i,j)}(x_{ij})_{\alpha_{1}, \alpha_{2}, \hdots \alpha_{z_{i,j}}} = \begin{cases}
         \left(\frac{c}{2}\right)^{\frac{1}{nL}}\exp(k_{i}\frac{x_{i,j}}{2^{j}} + \frac{a}{nL}) & \alpha_{1} = \alpha_{2} = \hdots = \alpha_{z_{i,j}} = 0 \\
         \left(\frac{c}{2}\right)^{\frac{1}{nL}}\exp(-k_{i}\frac{x_{i,j}}{2^{j}} - \frac{a}{nL}) & \alpha_{1} = \alpha_{2} = \hdots  = \alpha_{z_{i,j}} = 1 \\
         0 & {\rm Otherwise}.
     \end{cases}
 \end{equation}

\subsection*{Interpolative construction - Tensor cross interpolation}

\begin{figure}[t!]
    \centering
    \includegraphics[width = \columnwidth]{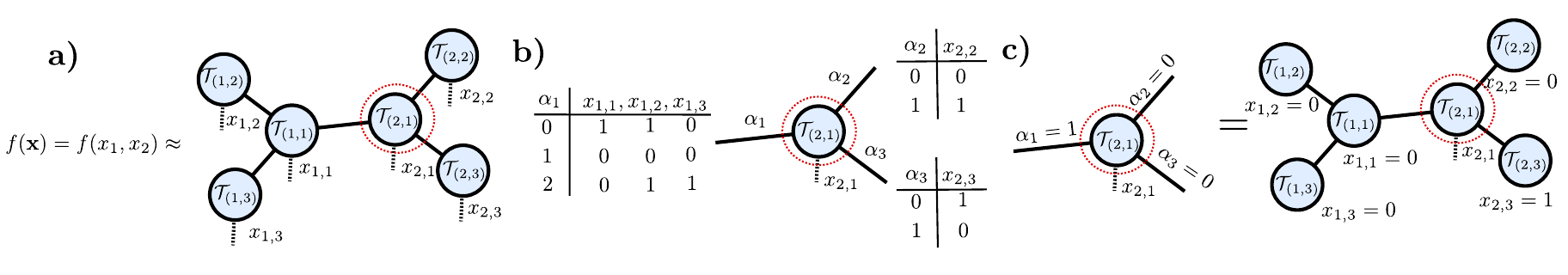}
    \caption{The interpolative gauge. \textbf{a)} Tree tensor network for a two dimensional function $f(\mathbf{x}) = f(x_{1}, x_{2})$ with the external indices decomposing the two continuous variables into binary strings of length three. \textbf{b)} The settings on the virtual indices $\alpha_{1}, \alpha_{2}, \alpha_{3}, \hdots$ of a local tensor map to configurations of the binary digits which the given edge connects to that local tensor. An example mapping is given by the tables shown. \textbf{c)} In the interpolative gauge, one of the tensors in the TTN has the property that their elements, for a given configuration of their virtual indices and external index, are equivalent to the contraction of the whole network for a specific setting of all the binary digits. 
    This information is encoded via the mapping between virtual indices and external ones. 
    The tensor cross interpolation (TCI) algorithm is an active learning algorithm which utilizes this gauge to move through the network, changing the gauge centre and updating neighboring pairs of local tensors in order minimize the infinity norm between their values $\mathcal{T}_{(i,j)}(x_{i,j})_{\boldsymbol{\alpha}}$ and some set of interpolation points of a target function $f(\mathbf{x})$.}
    \label{fig:InterpolativeGauge}
\end{figure}

The \emph{tensor cross interpolation} (TCI) algorithm, also known as TT-cross, computes or ``learns'' a tensor network $\mathcal{T}(\mathbf{x})$ which interpolates a ``target'' function $f(\mathbf{x})$ 
\cite{Oseledets2010V2,savostyanov2014quasioptimality,dolgov2020parallel,nunez2022learning,fernandez2024learning}. 
Assuming that the function can be computed efficiently for arbitrary inputs, the TCI algorithm  queries the function at adaptively determined points known as ``pivots'' 
to improve tensors in the network, attempting to minimize the infinity norm ${\rm sup}_{\mathbf{x}}(\vert f(\mathbf{x}) - \mathcal{T}(\mathbf{x}) \vert)$ while dynamically adapting the ranks of the network.
Though TCI is formally defined for functions of discrete variables $f(i_1, i_2, \ldots, i_n)$, it can be applied to continuum functions
by approximating continuous inputs $\mathbf{x}$ as binary index collections as in Sec.~\ref{sec:preliminaries}.

TCI is conventionally formulated for tensor trains (tensor networks with a one-dimensional topology) \cite{fernandez2024learning}, 
but here we generalize it to arbitrary tree networks. In our tree generalization of TCI, 
we define a tensor network gauge we call the ``interpolative gauge''. 
In this gauge, some specific local tensor $\mathcal{T}_{(i,j)}$, known as the ``center'' tensor, 
has the property that each of its elements corresponds exactly to an entry of the ``full'' tensor represented by the entire network 
(i.e.\ that would be formed by contracting all of the tensors in the network together). 
The other tensors in the network act to interpolate the values of the full tensor not contained in the center tensor.
Thus the interpolative gauge gives one access to or knowledge of certain values of the (exponentially large) full tensor through smaller, local
network tensors.  

Going into more detail, the center tensor has elements $\mathcal{T}_{(i,j)}(x_{i,j})_{\boldsymbol{\alpha}_{i,j}}$, 
where \newline $\boldsymbol{\alpha}_{i,j} = (\alpha_{1}, \alpha_{2}, ..., \alpha_{z_{i,j}})$ are special values of the virtual indices of $\mathcal{T}_{(i,j)}$.
In the interpolative gauge, additional ``pivot'' information is stored alongside these values saying how they map onto settings 
of the external indices. 
Figure~\ref{fig:InterpolativeGauge}a) shows a network whose center tensor is $\mathcal{T}_{(2,1)}$.
Pivot entries are shown for the virtual indices in Fig.~\ref{fig:InterpolativeGauge}b), for example setting $\alpha_1 = 0$ corresponds to setting $(x_{1,1},x_{1,2},x_{1,3}) = (1,1,0)$. In particular, as illustrated in Fig.~\ref{fig:InterpolativeGauge}(c), the elements $\mathcal{T}_{(2,1)}(x_{2,1})_{\boldsymbol{\alpha}_{2,1}=(1,0,0)}$ of the center tensor correspond exactly to the full tensor values $\mathcal{T}(0,0,0,x_{2,1},0,1)$.

On a tree tensor network, our tree TCI algorithm starts by making an initial guess for the tree tensor network and bringing it into the interpolative gauge.
Specifically, the interpolative gauge is obtained by first choosing some tensor to be the root, or gauge center, of the tree. The tensors corresponding to the leaves of the tree are then matricized and an interpolative decomposition of the resulting matrices is computed. The interpolative decomposition (ID) of a matrix $M$ is a factorization $M \approx C P^{-1} R = CZ$ such that the columns of $C$ are specific columns of the matrix $M$ and $Z = P^{-1}R$ interpolates any remaining columns of $M$ not contained in $C$. If the matrix $M$ has approximate rank $r$, then $C$ can be chosen to have slightly more than $r$ columns. 
For further discussion of computing ID factorizations with the fewest number of columns, see Refs.~\cite{fernandez2024learning, Fornace2024}. Here we use the prrLU (partial rank revealing LU) decomposition to explicitly implement the ID in a stable fashion, as is also done in Ref.~\cite{fernandez2024learning}.
After computing the ID of the leaf tensors, they are replaced by the $Z$ matrices and the $C$ matrices are contracted into the parent tensor towards the designated root. The algorithm continues by next computing the ID of these parent tensors and multiplying the $C$ tensors toward the root until all of the network consists of interpolating $Z$ tensors surrounding the root/ center tensor.

Once in the interpolative gauge, the interpolation quality can be improved by contracting the center tensor with a neighboring tensor. The resulting combined tensor denoted $\Pi$  no longer contains exact values of the full tensor, but is only an interpolation. This fact allows one to check point-wise how well the interpolation is matching the target function, and values which deviate by too much can be replaced by exact values obtained by calling the function. More efficient update strategies, which we do not use in this work, can be defined such as ``rook pivoting'' or ``block rook pivoting'' \cite{dolgov2020parallel,fernandez2024learning}.
After the update, an interpolative decomposition is performed on the $\Pi$ tensor to restore the interpolative gauge and move the gauge center to the neighboring location. The full algorithm proceeds by contracting the new center with another neighbor, updating, and so on until every bond of the tree is visited once, comprising a single full ``sweep'' of the algorithm. 

It is worth pointing out how this gauge relates to the tensor train constructions of Reference \cite{fernandez2024learning}. That work outlines the TCI algorithm for tensor trains by introducing the `TCI' form [Eq. (34) of Ref.~\cite{fernandez2024learning}], which keeps track of both a pivot tensor $P^{-1}$ on each of the bonds and a site tensor $T$ on each of the sites. These stem from keeping track of all three tensors from the ID $M \approx CP^{-1}R$ instead of absorbing the pivot $P^{-1}$ into $R$. Reference \cite{fernandez2024learning} also introduces, however, a `CI-canonical' form [Eq. (55) of Ref.~\cite{fernandez2024learning}], where a single site is designated as the gauge center and the pivot tensor on each bond is instead absorbed into the neighboring site tensor lying on the far side of that bond from the center [e.g. $B_{\ell} = P^{-1}_{\ell-1}T_{\ell}$ for sites to the right of the center, Eq. (54) of Ref.~\cite{fernandez2024learning}]. This leaves the center tensor holding exact values of the full tensor, with the remaining tensors acting to interpolate. For a tensor train, the interpolative gauge we utilize here is precisely this CI-canonical form, and we emphasize that the gauge itself is therefore not new. What we introduce is its extension to a generic tree, where the pivot tensors are retained on the outward side of every bond as the interpolating $Z$ tensors are swept in along each branch towards a chosen root. This extension follows rather naturally from the standard gauge center construction for tree tensor networks --- in the same sense that placing an orthogonality center on a tree via a sequence of SVD or QR decompositions from the leaves inwards is standard practice \cite{Schollwoeck2011, TindallGauging2023}.

\section{Numerical results for function construction}

\begin{figure}[t!]
    \centering
    \includegraphics[width = \columnwidth]{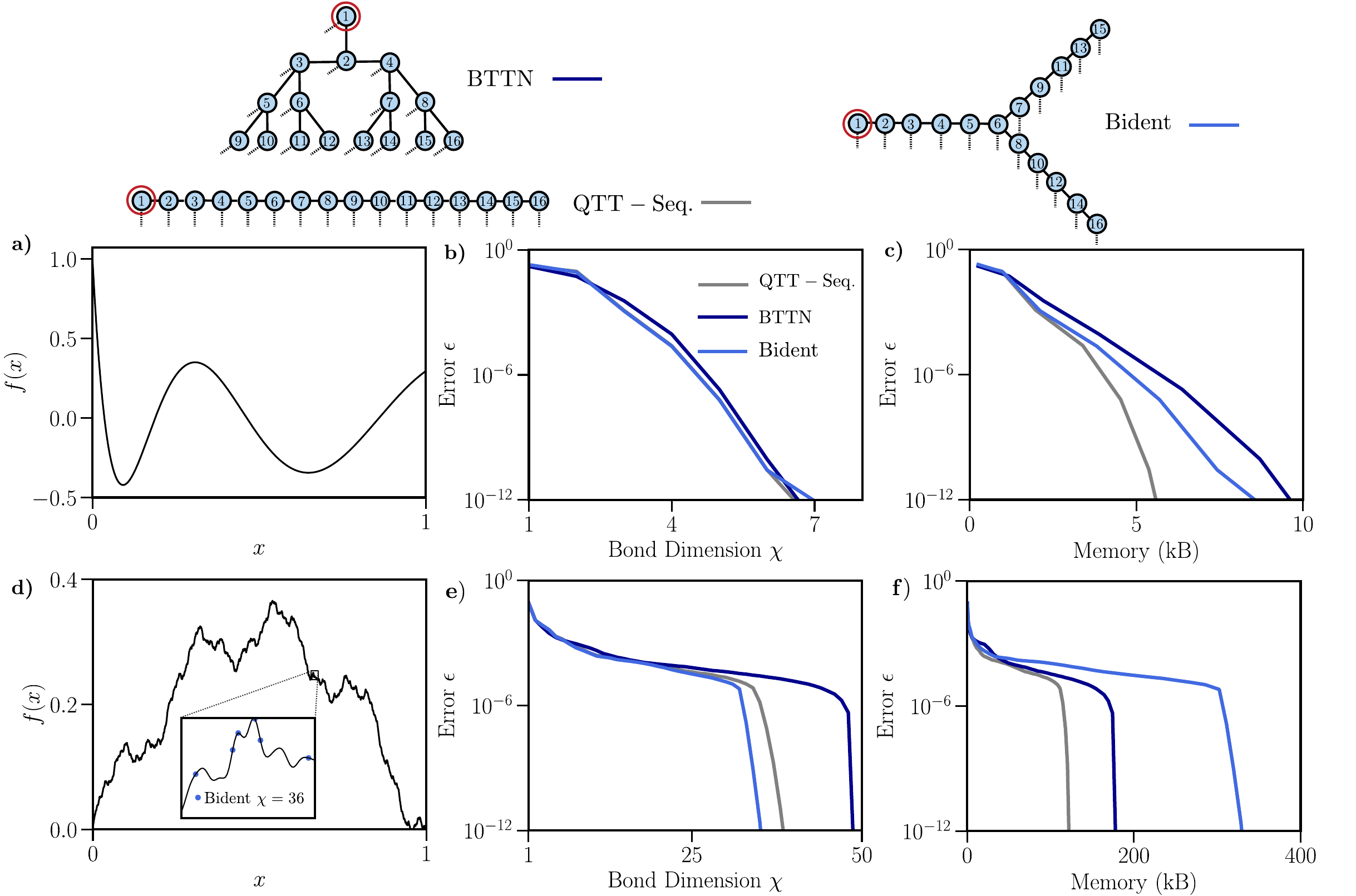}
    \caption{Comparison of different tree tensor networks, with $L = 16$ vertices, for compressing two different one-dimensional functions $f(x)$ with a binary decomposition $x = 0.x_{1}x_{2}...x_{16}$. Error $\epsilon$ is calculated via Eq.(\ref{Eq:Errors}), sampling the function over $10^{3}$ random grid points. The different labeled trees  used are shown at the top and the bits are numbered from most significant to least significant: with the most significant digit circled in red. Two functions are considered. Top plots) the Laguerre polynomial $f(x) = L_{n}(x) = \sum_{k = 0}^{n} \binom{n}{k}\frac{(-1)^{k}}{k!}x^{k}$ with $n = 40$ and Bottom plots) the Weierstrass function $f(x) = \sum_{k=1}^{n}\frac{\sin(\pi k^{a} x)}{\pi k^{a}}$ with $a = 3$ and $n = 25$.  Left panels: Sketch of the functions considered over $x \in [0,1]$. Inset of \textbf{d)} shows a zoomed-in region of the function with data points corresponding to the values obtained from the Bident tree tensor network with $\chi = 36$.  Middle panels:
    Error versus bond dimension of the tensor networks. Right panels: Error versus memory requirement for storing the tensor networks ---assuming $8$ bytes for a floating point number.
    }
    \label{fig:1DFunctionCompression}
\end{figure}

In this section we will compare the effectiveness of different tree tensor network topologies for representing a target function $f(\mathbf{x})$, using both direct methods and the tree tensor cross interpolation algorithm. The structure of the tensor network is specified by a labeled tree $\mathcal{T}$ and we will assess the effectiveness of such a tree by the error measures
\begin{equation}
    \epsilon = \frac{1}{\vert g \vert}\sum_{\mathbf{x} \in g}\vert f(\mathbf{x}) - \mathcal{T}(\mathbf{x}) \vert, \qquad \epsilon_{\infty} = \vert f(\mathbf{x}) - \mathcal{T}(\mathbf{x}) \vert_{\infty} = {\rm sup}_{\mathbf{x} \in g}\vert f(\mathbf{x}) - \mathcal{T}(\mathbf{x}) \vert
    \label{Eq:Errors}
\end{equation}
where $\mathcal{T}(\mathbf{x})$ corresponds to the contraction of the given tensor network for a specified configuration $\mathbf{x}$ of the external indices. Here, $g$ is a randomly chosen subset of the full $2^{nL}$ grid points which is taken to be sufficiently large to avoid any sampling bias. The same subset $g$ is used when comparing the efficacy of different TTNS for the same function $f(\mathbf{x})$. The measure $\epsilon$ can be utilized to identify on average how well the TTN approximates a function over the points in $g$ while $\epsilon_{\infty}$ is able to critically identify outliers and make guarantees on whether the function is well approximated for any point in $g$.

\begin{figure}[t!]
    \centering
    \includegraphics[width = \columnwidth]{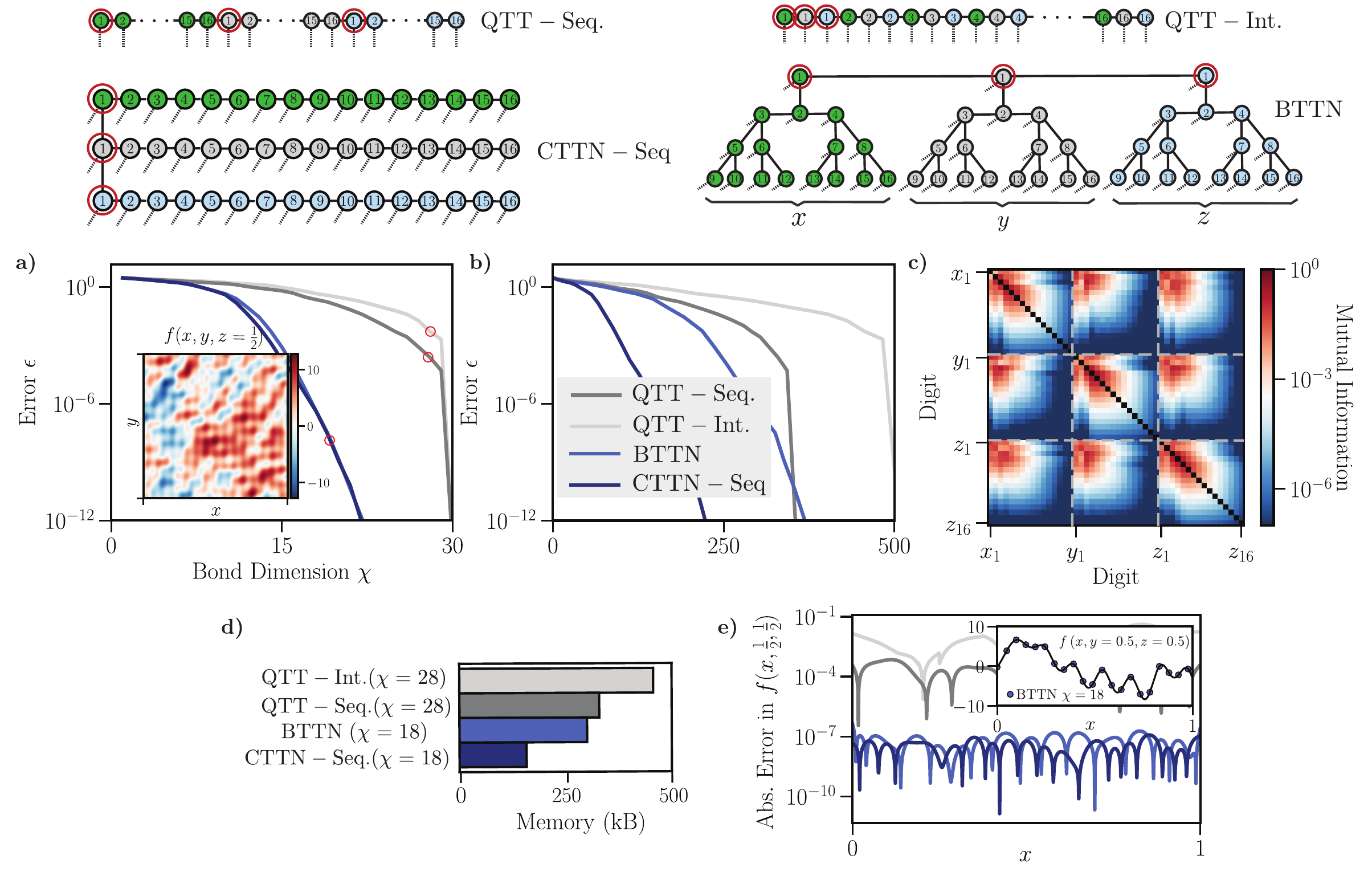}
    \caption{Comparison of different tree tensor networks for compressing the three dimensional function $f(\mathbf{r}) = \sum_{j=1}^{n}\cos(j \mathbf{k_{j}} \cdot \mathbf{r})$ with $n= 30$, $\mathbf{r} = (x,y,z)$ and $\mathbf{k_{j}} = (k^{x}_{j}, k^{y}_{j}, k^{z}_{j})$ with $k^{\alpha}_{j} \in \mathcal{N}(0,1)$. We use $L = 16$ bits per dimension. Error is calculated as $\epsilon$ via Eq.(\ref{Eq:Errors}), sampling the function over $10^{3}$ random grid points. Different trees used are shown at the top. Bits for $x$, $y$ and $z$ are colored in green, grey and light blue respectively and numbered sequentially from most significant to least significant.
    \textbf{a)} Error versus bond dimension. Inset shows a heatmap of the function at $z = \frac{1}{2}$. \textbf{b)} Error versus memory requirement for storing the tensor networks --- assuming $8$ bytes for a single floating point number.
    \textbf{c)} Mutual information matrix --- calculated via Eq. (\ref{eq:SMMutualInfo}) --- encoding the correlations between the binary digits. Calculated by sampling the function $10^{4}$ times to build up an approximate reduced density matrix for the given pair of bits. Each dashed block encodes the matrix of correlations between the bits of two given dimensions. \textbf{d)} Memory requirements for the specified TTNs at the given bond dimensions. \textbf{e)} Absolute error over a one-dimensional slice of the function for the bond dimensions in d). Inset (black line) shows the values of the function with corresponding results for the coupled binary tree tensor network at $\chi  =18$. 
    }
    \label{fig:3DFunctionCompression}
\end{figure}

\textit{1D Function Construction} - We consider two emblematic single-variable functions: a Laguerre polynomial and a truncated series representation of the Weierstrass function. We compare these functions represented on three different labeled trees in Fig.~\ref{fig:1DFunctionCompression}: a one-dimensional path (i.e. a QTT) with the most-to-least significant bits ordered from left to right, a binary tree with the more significant bits nearer the root, and a three-pronged tree with the most significant digit on the end of one of the prongs. 
We use our direct constructive methods to build an exact tensor network representation of a given target function $f(\mathbf{x})$ on the specified labeled tree and then compare the effect on the error $\epsilon$ when systematically truncating down the bond dimension of the network (via a singular value decomposition of local pairs of tensors) from $\chi = \chi_{\rm max}$ to $\chi = 1$.

For the Laguerre polynomial the function is continuous and smooth. We find, despite the high-order nature of the polynomial, that the function can be represented with an error of $\epsilon \sim 10^{-12}$ with maximum bond dimension $\chi = 7$ on any of the labeled trees. The tensor train with sequential digit ordering ($x_{1}, x_{2}, ..., x_{n})$ is slightly more effective in terms of error vs required memory to store the tensor network. 
The Weierstrass function, meanwhile, is a more complex function. The truncated series we consider is highly oscillatory, with frequencies growing as $k^{3}$. This complexity explains why a much larger bond dimension is required to exactly capture the finite-series realization of the function ($\chi \sim 40$). Here we find that the tensor train Ansatz with sequential digit ordering is noticeably more effective than the other labeled trees. The tensor train corresponds to a network with the lowest co-ordination number while still being connected. This directly translates into the lowest memory cost of $\chi^{2}$ for the tensors in the network. Moreover for typical one-dimensional functions the leading bits, and their correlations with each other, are the most important and so by clustering them all at the start of the train this allows the Ansatz to capture those fundamental correlations at a low cost.

\textit{3D Function Construction} - In Fig.~\ref{fig:3DFunctionCompression} we move on to consider a three-dimensional function which is the sum of $n = 30$ random plane waves of increasing frequency and compare four topologies: two tensor trains with different, commonly used, digit orderings (interleaved and sequential), a tree consisting of three separate binary trees, one for each dimension, coupled together at their roots (where the most significant bits are placed) and a comb tree  consisting of three sequential tensor trains, one for each dimension, coupled at the first digit. 
Here we find that the coupled binary tree tensor network (BTTN) and the comb tree tensor network (CTTN) are both better Ansatzes for the function at hand and compress much more effectively under truncation of the internal bonds. The CTTN performs particularly well: For a fixed memory cost it can achieve orders of magnitude lower error than the trains. For instance, it is able to represent the function with an error $\epsilon \sim 10^{-6}$ with a bond dimension of $\chi = 15$ while the tensor trains each require a bond dimension of $\chi = 30$ to reach such an error. As $n = 30$ the function can be represented exactly (as a sum of 30 complex exponentials, taking the real part at the end) on any network with a bond dimension $\chi = 30$ and thus this shows that the tensor trains are an ineffective representation.  In Fig. \ref{fig:3DFunctionCompression}e) we show the absolute error over a one-dimensional slice of the function. The more structured trees (CTTN and BTTN) are seen to achieve much lower errors at a lower memory cost.

To support our analysis, we also compute the correlation measure $M(x_{A}, x_{B})$ between two binary digits $x_{A} = x_{i,j}$ and $x_{B} = x_{i',j'}$ for a given function $f(\mathbf{x})$ by interpreting the function values as coefficients of a wavefunction and computing the quantum mutual information $M(x_{A}, x_{B})$ \cite{Carr2017} by building an approximate representation of the two-body reduced density matrix $\rho_{A,B}$ via sampling of the function (see Supplementary Material for calculation details). The value of $M(x_{A}, x_{B})$ is a good proxy for the correlations between the two bits and the larger this value the closer bits $x_{A}$ and $x_{B}$ will need to be in the tensor network in order to accurately encode their correlations with a fixed bond dimension. The plot of the mutual information matrix in Fig.~\ref{fig:3DFunctionCompression}c shows that the most significant binary digits are strongly correlated with the other significant bits within their dimension and with those in other dimensions. Meanwhile, the least significant bits are typically correlated only with bits within their respective dimension. These lead us to understand why the CTTN is far more effective: it keeps the more significant bits in a given dimension clustered together and close to their counterparts in other dimensions. The BTTN also achieves this, but has some of the less significant digits spatially separated from each other, which likely explains its worse performance compared to the CTTN.
We emphasize that our results here are not specific to the random frequencies and amplitudes chosen: we observed the same qualitative results for any random realization of the plane wave frequencies.   

\begin{figure}[t!]
    \centering
    \includegraphics[width = \columnwidth]{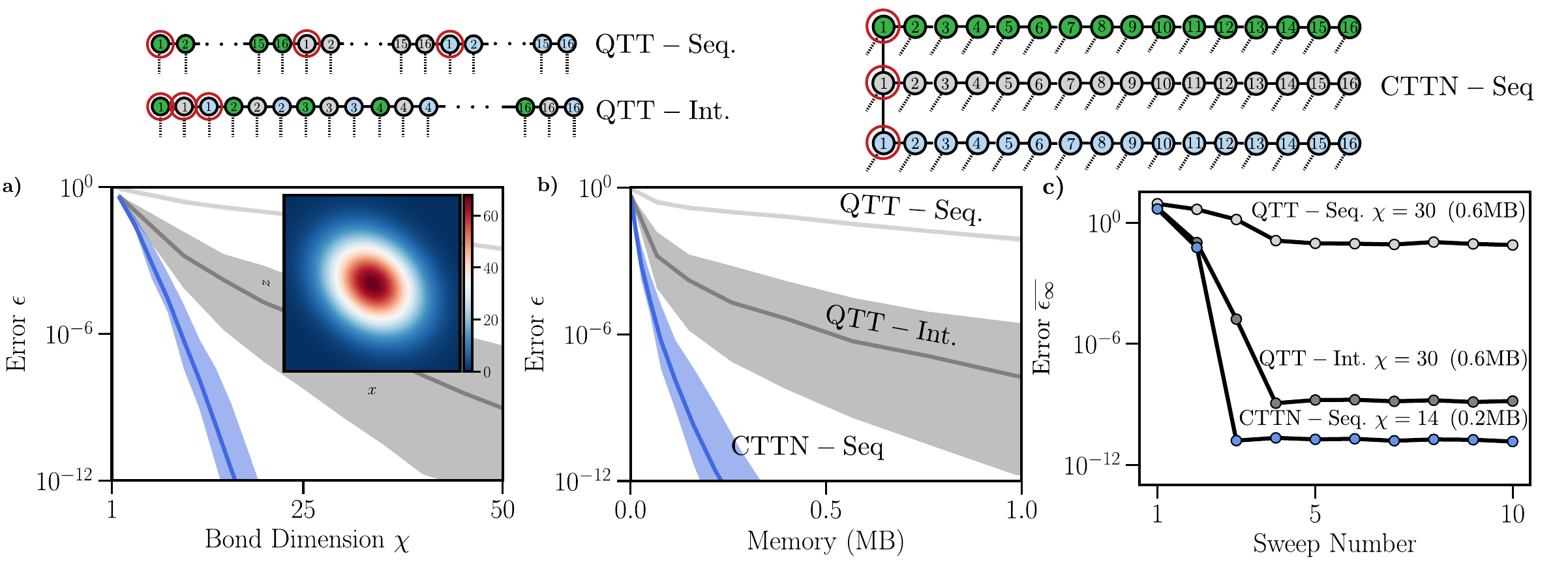}
    \caption{Comparison of the effectiveness of different tree tensor networks with $L = 16$ bits per dimension for learning the multinormal probability density function $f(\mathbf{r}) \propto \exp( - ((\mathbf{r} - \mathbf{\mu})^{T} M^{-1} (\mathbf{r} - \mathbf{\mu}))$ via the tensor cross interpolation algorithm. Here $M$ is an $n \times n$ covariance matrix and $\mathbf{\mu} = (\mu_{1}, \mu_{2}, ..., \mu_{n})$ is the mean vector. We consider $n =3$ with $\mathbf{r} = (x,y,z) \in [0,10)^{3}$ and $\mathbf{\mu} = (5,5,5)$. Results are obtained from drawing $N = 10$ instances of $M$ ($M_{1}, M_{2}, \hdots M_{10}$) from the LKJ distribution \cite{Lewandoski2009} with shape parameter $\eta = 50$. Different trees used are shown at the top. Bits for $x$, $y$ and $z$ are colored in green, grey and light blue respectively. The most significant digit in each dimension is circled in red. 
    \textbf{a-b)} Error $\epsilon$, calculated via Eq. (\ref{Eq:Errors}) after $10$ sweeps of the TCI algorithm, versus bond dimension and memory cost for the tensor networks. 
    The solid lines shows the mode of the error over the $10$ realizations of $M$ while the shaded area shows the range of the error, i.e. for any $M$ sampled the error $\epsilon$ for a given bond dimension. Inset) shows a heatmap of the function over $(x,z) \in [3,7]^{2}$ with $y = 5$.
    \textbf{c)} Average value for the infinity norm $\epsilon_{\infty}$ (see Eq. (\ref{Eq:Errors})) over a given sweep of the TCI algorithm for $M = M_{1}$ and the three tensor networks at the specified bond dimensions and memory cost. 
    }
    \label{fig:TCI3D}
\end{figure}

\textit{TCI Function Construction} -
In Fig.~\ref{fig:TCI3D} we use the TCI algorithm to build representations of the trivariate Gaussian probability density function $f(\mathbf{r}) \propto \exp( - (\mathbf{r} - \mathbf{\mu})^{T} M^{-1} (\mathbf{r} - \mathbf{\mu}))$ where $\mathbf{r} = (x,y,z)$, $\mathbf{\mu} = (\mu_{x}, \mu_{y}, \mu_{z})$ is the mean vector, and $M$ is a covariance matrix which we sample from the Lewandowski-Kurowicka-Joe (LKJ) distribution with shape parameter $\eta = 50$ \cite{Lewandoski2009}, which controls the weight of the off-diagonal correlations.
Beyond its statistical significance, this function physically represents the ground-state density of three coupled quantum harmonic oscillators, where the shape parameter $\eta$ directly controls the spatial coupling between the coordinates. Such coupled systems also serve as standard toy models in continuous-variable quantum computing \cite{Weedbrook2012}.

We compare results from the TCI algorithm when varying the maximum allowed bond dimension for different tree tensor networks: two tensor trains with commonly used digit orderings (sequential and interleaved) for multi-dimensional functions and a comb tree tensor network consisting of sequentially ordered tensor trains  (CTTN - Seq) coupled by their most significant digit.

Similarly to our results for random plane waves we find that for all covariance matrices we sample, the comb tree systematically outperforms the tensor trains. Firstly, for a fixed memory cost it achieves orders of magnitude lower error $\epsilon$ than the tensor trains (see Fig. \ref{fig:TCI3D}b). Moreover, the variance in the error vs memory curves is much lower than for the quantics tensor train with an interleaved ordering, indicating it is a much more effective and consistent Ansatz. The variance for the QTT with sequential ordering is also very low, but the errors are drastically worse than the other two Ansatzes (we are unable to converge the error to below $\epsilon \sim 10^{-2}$ for any of the covariance matrices sampled). This low variance is therefore just an indicator that the Ansatz is consistently poor. 

We emphasize that the effectiveness of the comb tree in comparison to the tensor trains is not at all specific to the shape parameter $\eta$ we chose. 
In the Supplemental Material we also show results for $\eta = 1$, which is equivalent to sampling uniformly from the space of all covariance matrices. For any given $M$ the structured tree offers a significantly better Ansatz than the tensor trains: with many orders of magnitude lower errors for a given memory cost. Due to the lower value of $\eta$, however, the variance (i.e. the size of the shaded area in Fig. \ref{fig:TCI3D}) in the bond dimension / memory required to accurately represent the function is much higher because certain matrices are sampled which have very significant correlations between the continuous variables. This makes the function harder to represent with a TTN Ansatz. In the Supplemental Material we also include walltime data for our TCI implementation, finding the comb tree tensor network is able to achieve a lower error than the tensor trains at a fixed walltime.

\section{Solving Fredholm equations with tree tensor networks}\label{sec:Fredholm}

Integral equations arise in many different scientific domains \cite{Jerri1999}. In physics, for instance, Fredholm integral equations of the second kind are ubiquitous in modelling phenomena such as quantum scattering \cite{Lippmann1950} and radiative transfer problems \cite{Domke1978}.  Analytical solutions of such equations are typically difficult to find and thus numerical methods are vital \cite{Guoqiang2001, Borzabadi2009, Kazemi2019}. Here we use the methods introduced in this paper to define an iterative tree tensor network (TTN) based numerical algorithm for Fredholm integral equations of the second kind. The solution is represented as a TTN and following the methods introduced in the paper, there is complete flexibility over the structure of the tree chosen.

We focus on the following Fredholm equations of the second kind 
\begin{equation}
    f(\mathbf{x}) = g(\mathbf{x})+  \lambda \int_{\mathbf{t} \in [0,1)^{n}} K(\mathbf{x},\mathbf{t}) f^{\alpha}(\mathbf{t}) d\mathbf{t} \qquad \alpha \in \mathbb{N},
    \label{Eq:Fredholm}
\end{equation}
where $\mathbf{x} = (x_{1},x_{2}, \hdots x_{n}) \in [0,1)^{n}$ and $\mathbf{t} = (t_{1},t_{2}, \hdots t_{n}) \in [0,1)^{n}$.
The integral kernel is $K(\mathbf{x},\mathbf{t}): [0,1)^{n} \times [0,1)^{n} \rightarrow \mathbb{R}$ and $g(\mathbf{x}): [0,1)^{n} \rightarrow \mathbb{R}$ is a given function. We wish to find the solution $f(\mathbf{x}): [0,1)^{n} \rightarrow \mathbb{R}$, setting $\lambda=1$ without loss of generality. In our examples we will focus on the non-linear case ($\alpha > 1$), however our algorithms and analysis also apply straightforwardly to the linear case ($\alpha=1$) as well.

\begin{figure}[t!]
    \centering
    \includegraphics[width = \columnwidth]{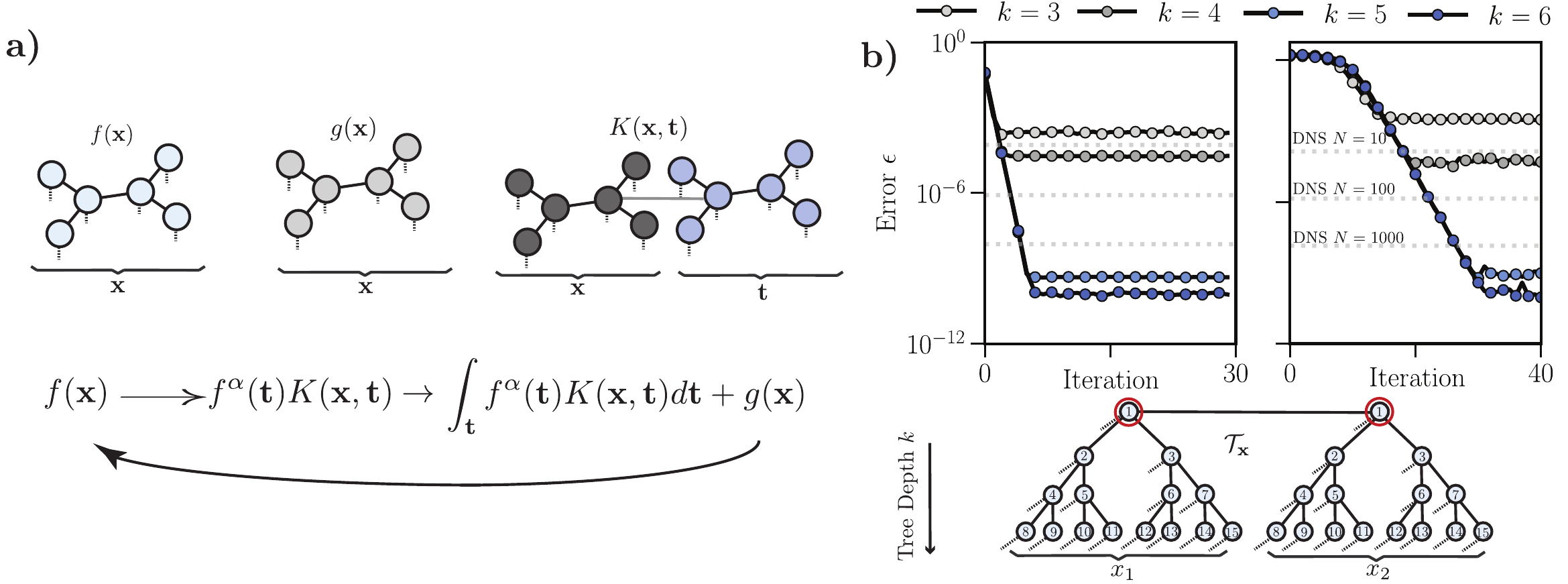}
    \caption{Tree tensor network method for solving (non-linear) Fredholm equations of the second kind $f(\mathbf{x}) = g(\mathbf{x}) + \int_{\mathbf{t}}f^{\alpha}(\mathbf{t})K(\mathbf{x}, \mathbf{t})d\mathbf{t}$.  \textbf{a)} An initial guess for the target function $f(\mathbf{x})$ is constructed as a tensor network over a labeled tree of depths $k=3,4,5,6$ with $L = 7,15,31,63$ bits per dimension respectively. 
    The additive function $g(\mathbf{x})$ is constructed as a tree tensor network of the same structure. The kernel $K(\mathbf{x}, \mathbf{t})$ is then constructed over as a tree tensor network formed from two copies of the tree, with one encoding the `real' variables $\mathbf{x}$ and the other the auxiliary variables $\mathbf{t}$. A single edge is also added between the two trees. The Fredholm equation can then be solved by the iterative steps expressed at the bottom and outlined in Alg. \ref{alg:Fredholm}. \textbf{b)} Results for two different Fredholm equations with a two dimensional solution $f(\mathbf{x}) = f(x_{1}, x_{2})$. Explicit expressions for $K(x_{1}, x_{2}, t_{1}, t_{2})$ and $g(x_{1}, x_{2})$ are given in the main text. The structure of the underlying tree for $f(x)$ is shown at the bottom with the binary digits numbered by their significance: the most significant digit in each dimension is circled in red. Error $\epsilon$ is calculated as in Eq. (\ref{Eq:Errors}), sampling over $100$ random grid points and comparing to the values for the exact solution, versus iteration of the solver. Increasing $k$ decreases the grid spacing super exponentially and thus drastically improves the error  until one starts to reach machine precision.
    Dashed lines show the converged error from the Direct Numerical Simulation (DNS) results of Ref.~\cite{Kazemi2019} with $N = 10, 100$ and $1000$ grid points. Left plot) Results for example 1 where the exact solution is $f(x_{1}, x_{2}) = \sin(x_{2})$. Right plot) Results for example 2 where the exact solution is $f(x_{1}, x_{2}) = 1/(1+x_{1} + x_{2})^{2}$.
    }
    \label{fig:FredholmResults}
\end{figure}

We consider a tree tensor network defined over a labeled tree $\mathcal{T}_{\mathbf{x}}$ as the Ansatz for $f(\mathbf{x})$ and perform the iterative procedure ($N$ being the number of iterations) illustrated in Fig.~\ref{fig:FredholmResults}) to attempt to solve Eq.~(\ref{Eq:Fredholm}). The procedure is also given in ``pseudo code'' in Algorithm~\ref{alg:Fredholm}. 
The kernel $K(\mathbf{x}, \mathbf{t})$ is constructed over the tree $\mathcal{T}_{\mathbf{x}} \cup \mathcal{T}_{\mathbf{t}}$, where $\mathcal{T}_{\mathbf{t}}$ is identical to $\mathcal{T}_{\mathbf{x}}$ except the vertices have been relabeled $x_{(i,j)} \rightarrow t_{(i,j)}$. A single edge $e_{\mathbf{x} \leftrightarrow \mathbf{t}}$ is added to $\mathcal{T}_{\mathbf{x}} \cup \mathcal{T}_{\mathbf{t}}$ between one of the binary digits in $\mathbf{x}$ and one in $\mathbf{t}$. One can view such a TTN construction of the kernel as the finite-rank decomposition $K(\mathbf{x},\mathbf{t}) = \sum_{i=1}^{\rm dim(\alpha_{e_{\mathbf{x} \leftrightarrow \mathbf{t}}})}u^{(i)}(\mathbf{x})v^{(i)}(\mathbf{t})$ where the function $u^{(i)}(\mathbf{x})$ is represented with a TTN with structure specified by $\mathcal{T}_{\mathbf{x}}$ and $v^{(i)}(\mathbf{t})$ is represented with a TTN with identical structure specified by $\mathcal{T}_{\mathbf{t}}$. Importantly, our generic TCI algorithm provides us with a method to approximately identify such a decomposition.

The partial integration of $f^{\alpha}(\mathbf{t})K(\mathbf{x}, \mathbf{t})$ over $\mathbf{t}$ is performed by multiplying the tensors with external indices corresponding to bits in $\mathbf{t}$ with the vector $T_{i,j} = (\frac{1}{2}, \frac{1}{2})$ and then contracting away those tensors. This process can be written diagrammatically as:
\begin{equation*}
    \adjincludegraphics[valign=c, width = 0.9\textwidth]{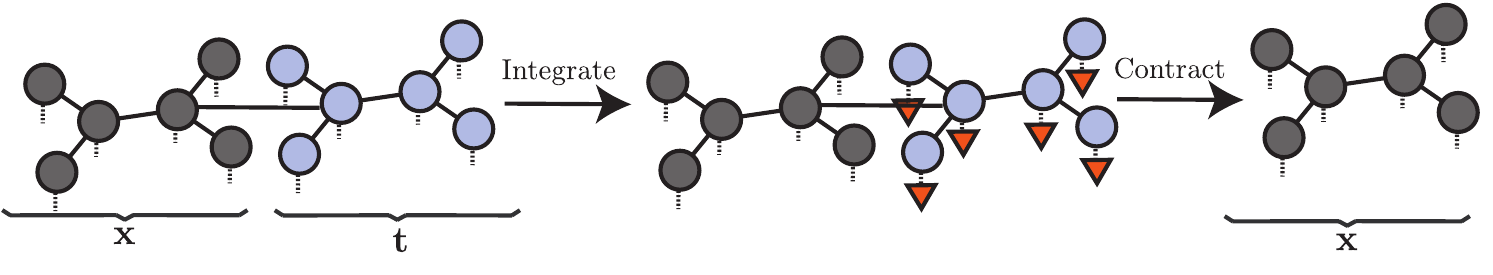},
    \label{Eq:PartialIntegration}
\end{equation*}

\begin{algorithm}[t!] 

\caption{Tree tensor network method for solving the non-linear Fredholm equation --- see Eq. (\ref{Eq:Fredholm})} \label{alg:Fredholm}
\begin{algorithmic}
\State Given (by using the TTN construction methods specified in Sec. \ref{sec:Construction}): Tree tensor network (TTN) representing the initial guess $f_{1}(\mathbf{x})$ with structure specified by the labeled tree $\mathcal{T}_{\mathbf{x}}$. TTN representing $g(\mathbf{x})$ with structure given by  $\mathcal{T}_{\mathbf{x}}$. TTN representing $K(\mathbf{x}, \mathbf{t})$ with structure represented by $\mathcal{T}_{\mathbf{x}} \cup \mathcal{T}_{\mathbf{t}}$. Where $\mathcal{T}_{\mathbf{t}}$ is a copy of $\mathcal{T}_{\mathbf{x}}$ but with the external indices mapped: $x_{(i,j)} \rightarrow t_{(i,j)}$.  There is \textit{one} additional edge between $\mathcal{T}_{\mathbf{x}}$ and $\mathcal{T}_{\mathbf{t}}$ in $\mathcal{T}_{\mathbf{x}} \cup \mathcal{T}_{\mathbf{t}}$. 
\For{$m$ in $1:{\rm N}$}
    \State Remap Variables:  $f_m(\mathbf{t}) \gets f_m(\mathbf{x})$ \Comment{The mapping is achieved via the index relabelling $x_{i,j} \rightarrow t_{i,j}$.}
    \State Multiply (Sec.~\ref{sec:Construction}): $f_m(\mathbf{x},\mathbf{t}) \gets K(\mathbf{x}, \mathbf{t}) f^\alpha_{m}(\mathbf{t})$
    \State Integrate (Sec.~\ref{sec:Fredholm}): $f_m(\mathbf{x}) \gets   f_{m}(\mathbf{x}, \mathbf{t})\prod_{i=1}^{n}\prod_{j=1}^{L}T_{i,j}(t_{i, j})$ \Comment{$T_{i,j}(t_{i, j}) = \frac{1}{2} \ \forall i, j$.}
    \State Add (Sec.~\ref{sec:Construction}): $f_{m+1}(\mathbf{x}) \gets f_m(\mathbf{x})+g(\mathbf{x}) $
\EndFor
\end{algorithmic}
\end{algorithm}

Importantly, the bond dimension of the tree tensor network at the end of each iteration, independent of the initial state, is guaranteed to be $\chi = \chi_{g} + \chi_{K}$ where $\chi_{g}$ is the bond dimension of $g(\mathbf{x})$ and $\chi_{K}$ is the bond dimension of the kernel $K(\mathbf{x}, \mathbf{t})$. Thus the success of the algorithm relies on finding accurate, low bond dimension representations of $K(\mathbf{x}, \mathbf{t})$ and $g(\mathbf{x})$ on the given labeled tree. This is because the algorithm's time complexity scales, at worst, as $\mathcal{O}(NnLr\chi_{K}^{z}(\chi_{g} + \chi_{K})^{\alpha z})$ --- with $z$ the co-ordination number of the tree $\mathcal{T}_{\mathbf{x}}$ and $r = {\rm dim}(\alpha_{e_{\mathbf{x} \leftrightarrow \mathbf{t}}})$ the rank of the kernel --- while the error on the integration scales as $\mathcal{O}(\exp(-L))$. Thus the complexity of the algorithm is based on the representation of $K(\mathbf{x}, \mathbf{t})$ and $g(\mathbf{x})$ and how convergent the fixed point procedure is, as opposed to the integration itself: which can be a limitation of DNS (Direct Numerical Simulation) solvers such as that outlined in Ref. \cite{Kazemi2019}.  We emphasize that the scaling we derive stems from the multiplication step in Algorithm \ref{alg:Fredholm}, which is by far the most expensive step of the algorithm. Methods for accelerating the multiplication of two TTNs would substantially benefit the runtime of this algorithm.

In Fig.~\ref{fig:FredholmResults} we present results from this method for two example non-linear Fredholm equations, with known two-dimensional solutions \cite{Kazemi2019}. In both examples we take $\mathcal{T}_{\mathbf{x}}$ to be a tree formed from a pair (one for each dimension) of binary trees of depth $k$ coupled by their roots (see Fig. \ref{fig:FredholmResults} for more details) giving $z = 3$. The more significant bits are placed nearer the roots of the tree. 
The examples we use correspond to
\begin{itemize}
    \itemsep-0.25em
    \item Example I: $K(x_{1}, x_{2}, t_{1}, t_{2})$ = $x_{1} x_2^2 t_{1} / 6$, $\alpha=3$ and $g(x_{1},x_{2}) = \sin(x_2)-c x_1 x_2^2$ with $c=(1-\cos(1)(\sin^2(1)/2+1))/18$. The exact solution is $f(x_{1}, x_{2}) = \sin(x_{2})$.
    \item Example II: $K(x_{1},x_{2},t_{1}, t_{2}) = \frac{x_{1}(1+t_{1} + t_{2})}{1+x_{2}}$, $\alpha=2$ and \[ g(x_{1},x_{2}) = 1/(1+x_1+x_2)^2-x_1/(1+x_2)/6\,. \] The exact solution is $f(x_{1}, x_{2}) = 1/(1+x_{1} + x_{2})^{2}$.
\end{itemize}
In the first example all of the relevant functions can be constructed exactly using our direct construction method for polynomials, giving $\chi_{g}, \chi_{K} = 6, 4$. In the second we are able to use TCI to construct $g$ and $K$ on the trees $\mathcal{T}_{\mathbf{x}}$ and $\mathcal{T}_{\mathbf{x}} \cup \mathcal{T}_{\mathbf{t}}$ with errors on the order of the grid spacing $\epsilon \sim O(2^{-L})$ while using $\chi_{g}, \chi_{K} = 10, 10$. These bond dimensions are constant with system size and the structure of $\mathcal{T}_{\mathbf{x}}$ is shown in Fig. ~\ref{fig:FredholmResults}. The rank of the Kernel is $r = 1$ in both cases.

We show the results of the solver in Fig.~\ref{fig:FredholmResults}b starting from a tensor network of bond dimension $\chi = 1$ representing the constant function $f_{1}(x_1,x_2) = 1$. In each example, we observe convergence of our solution to a given error $\epsilon$ (in comparison to the exact solution) which is controlled by the number of bits $L$ which we take to decompose each continuous variable. Notably, the errors we achieve are always on the order of the grid spacing $\epsilon \sim O(2^{-L})$, which is the error in our representation of $K$ and $g$ and our integration technique. For sufficient $L$ this exponential precision allows us to reach a much higher accuracy than the quadrature-based DNS methods benchmarked in Ref.~\cite{Kazemi2019}.

\section{Conclusion}
In this paper we have introduced the tree tensor network (TTN) Ansatz for representing functions and solving problems in continuous space, generalizing beyond the almost-exclusively used one-dimensional tensor train Ansatz. We provided direct and indirect (via an extension of the tensor cross interpolation algorithm) methods for constructing tree tensor network (TTN) representations of mathematical functions. We identified a direct construction of polynomial functions, with an upper bound on the maximum bond dimension in the tree which is independent of the network topology. For multi-dimensional functions we find that TTNs with more complex structure --- such as comb tree tensor network (CTTN) and coupled binary tree tensor networks (BTTN) --- can be a much more effective Ansatz than the (quantics) tensor train. This is because these more structured TTNs can simultaneously keep the leading binary digits close to their counterparts within the same dimension and in other dimensions. This efficacy appears to be most apparent for the CTTN which maintains a sequential order amongst all the digits in each dimension, while keeping the leading digits in each dimension close together. We believe this TTN will therefore make a very effective Ansatz for multivariate function compression  going forward.

Using the tools introduced in this paper we also introduced a new iterative TTN-based solver for non-linear Fredholm integral equations. For our algorithm, provided the iterative solver converges, the bond dimension of the solution is therefore boundable in terms of the bond dimension of the kernel. The algorithm's success rests on finding an effective compressed representation of the Kernel and the convergence properties of the fixed point equation --- versus any limitation in the precision of the integration or grid used, which can affect more direct numerical solvers \cite{Kazemi2019}.

It is worth adding one caveat, however, on the practical benefit of these more structured trees. Throughout this work we have quantified compression by the memory required to reach a given accuracy, and this does not automatically translate into runtime: the cost of manipulating a tensor grows as a power of the bond dimension which itself increases with its co-ordination number $z$, with the interpolative decomposition scaling as $\chi^{4}$ for our comb tree versus $\chi^{3}$ for a tensor train. In the example we time in the Supplemental Material the comb tree nonetheless reaches a given error in less walltime than the trains, but this need not hold in general --- related cost comparisons for trains and trees representing quantum states which obey an area law are analyzed in Ref.~\cite{Barthel2026} --- and the trade-off is thus worth assessing for the specific function and algorithm at hand.

Looking forward, an important outstanding question is identifying a relevant cost function, and a search algorithm for finding the tree corresponding to its extrema, which allows one to identify good candidates for the correct tree structure for representing a given function. Potential heuristics could include those based on the quantum mutual information which have had success in determining the optimal ordering for sites in matrix product states when applied to quantum chemistry problems \cite{Rissler2006, Barcza2011, Ali2021}. 

Finally, we wish to emphasize that the generality of the work described here means that a wide range of TTN-based numerical algorithms can be built, significantly broadening the scope and potential of tensor-network based numerical methods. For problems where the solution is a multivariate function with significant inter-dimensional correlations, our results suggest moving away from the tensor train Ansatz and working with more structured tree tensor networks could push the state-of-the-art for problem solving. Such functions, for instance, arise in turbulent solutions of the Navier-Stokes equation and are currently pushing the limits of the tensor train Ansatz \cite{Jaksch2023, gourianov2024}.

\section*{Acknowledgments}
We would like to thank Matthew Fishman, Ryan Anselm and Marc Ritter for useful discussions.
The tools and methods described in this paper are contained within the ITensorNumericalAnalysis.jl library \cite{ITensorNumericalAnalysis}: an open-source Julia package built on top of  ITensors.jl \cite{itensor-r0.3} and \textit{TensorNetworkQuantumSimulator.jl} \cite{rudolph2025simulatingsamplingquantumcircuits} for constructing and optimizing tree tensor network representations of continuous functions, with the tree structure freely definable by the user.

\paragraph{Funding information} The authors are grateful for ongoing support through the Flatiron Institute, a division of the Simons Foundation.

\clearpage
\begin{appendix}
\numberwithin{equation}{section}

\section{Appendix}
\subsection*{Proof the TTN construction in Sec \ref{sec:Construction} contracts to the polynomial \newline $f(x) = \sum_{k = 0}^{d}c_{k}x^{k}$.}
Here we prove that a tree tensor network whose local tensors have elements specified by Eq. (\ref{Eq:PTensors}) will contract down to the polynomial $f(x) = \sum_{k = 0}^{d}c_{k}x^{k}$. Throughout this appendix we write $x_{j}$ for the weighted bit $x_{j}/2^{j}$, so that, e.g., the first line of Eq.~(\ref{Eq:PTensors}) reads $\mathcal{T}_{j}(x_{j})_{\beta} = x_{j}^{\beta}$. First, we observe that the elements of the local tensors satisfy the following properties, 
\begin{align}
    &\sum_{\alpha_{z_{j}-1} = 0}^{d}(x_{j'})^{\alpha_{z_{j}-1}}\mathcal{T}_{j}(x_{j})_{\alpha_{1}, \alpha_{2}, \hdots, \alpha_{z_{j}-1}, \beta} = C_{\alpha_{1}, \alpha_{2}, \hdots, \alpha_{z_{j}-2}, \beta} (x_{j} + x_{j'})^{f_{\alpha_{1}, \hdots, \alpha_{z_{j}-2}, \beta}} \\ = \qquad &\mathcal{T}_{(j, j')}(x_{j} + x_{j'})_{\alpha_{1}, \alpha_{2}, \hdots, \alpha_{z_{j} - 2}, \beta} \notag \qquad z_{j} > 2, \notag \\
    &\sum_{\alpha_{1} = 0}^{d}(x_{j'})^{\alpha_{1}}\mathcal{T}_{j}(x_{j})_{\alpha_{1}, \beta} = (x_{j} + x_{j'})^{\beta} = \mathcal{T}_{(j, j')}(x_{j} + x_{j'})_{\beta}  \qquad z_{j} = 2.
    \label{Eq:PTensorProperty}
\end{align}
This means that the contraction $\mathcal{T}_{j} \cdot \mathcal{T}_{j'}$ of one of the tensors on the leaves of the tree $\mathcal{T}_{j'}$ and its parent $\mathcal{T}_{j}$ yields a new tensor $\mathcal{T}_{(j,j')}$ (with two external indices corresponding to $x_{j}$ and $x_{j'}$ and $z_{j,j'} = z_{j} + z_{j'} - 2 = z_{j} - 1$ virtual indices corresponding to the set difference of the virtual indices of $\mathcal{T}_{j}$ and $\mathcal{T}_{j'}$) whose properties are still completely specified by Eq. (\ref{Eq:PTensors}).

We can utilize this property to iteratively prove the contraction of the tree yields $f(x)$. Following Eq. (\ref{Eq:PTensorProperty}) the binary digits effectively sum under contraction of the tensors on the leaves of the tree with their parents. This means we can repeat the process of contracting the leaves of the tree onto their parents until we are left with the root tensor $\tilde{\mathcal{T}}_{r}$ surrounded by $z_{r}$ tensors each with a single virtual index $\beta$ whose elements are specified by $(z_{k})^{\beta} = (\sum_{x_{j} \in {\rm Branch}(\beta)}x_{j})^{\beta}$ where the sum runs over all of the bits $x_{j}$ which the branch specified by the virtual index $\beta$ connects to $x_{r}$. 
The following property holds for the elements of $\tilde{\mathcal{T}}_{r}$
\begin{equation}
        \sum_{\alpha_{1} = 0}^{d}(z_{1})^{\alpha_{1}}\sum_{\alpha_{2} = 0}^{d}(z_{2})^{\alpha_{2}}\hdots \sum_{\alpha_{z_{r}} = 0}^{d}(z_{r})^{\alpha_{z_{r}}}
    \tilde{\mathcal{T}}_{r}(x_{r})_{\alpha_{1}, \alpha_{2}, \hdots, \alpha_{z_{r}}} = \sum_{k = 0}^{d}c_{k}(z_{1} + z_{2} + z_{3} + \hdots z_{r} + x_{r})^{k} \qquad z_{i} \in \mathbb{C}
\end{equation}
and so we can contract these satellite tensors onto $\tilde{\mathcal{T}}_{r}(x_{r})$ and arrive at \newline
$f(x) = \sum_{k=0}^{d}c_{k}( \sum_{j=1}^{L}x_{j} )^{k} = \sum_{k=0}^{d}c_{k}x^{k}$. The proof is complete.

\textit{Elevating to higher dimensions} - We can also represent polynomials as tree tensor networks in the case there are multiple continuous variables: i.e. we wish to build a TTN representation of $f(\mathbf{x}) = f(x_{1}, x_{2}, \hdots x_{n}) = f(x) = \sum_{k = 0}^{d}c_{k}x^{k}$, where $x \in \{x_{1}, x_{2}, \hdots x_{n}\}$. The construction is achieved as before, with the tensors $\mathcal{T}_{(j)}$ with an external index corresponding to $x$, defined via Eq. (\ref{Eq:PTensors}). For the tensors $\mathcal{T}_{(i,j)}$ where $x_{i} \neq x$ their elements are given as
\begin{align}
    &\mathcal{T}_{(i,j)}(x_{i,j})_{\alpha_{1},\alpha_{2},\hdots,\alpha_{z_{i,j}-1}, \beta} = \begin{cases}
        1 & \sum_{l=1}^{z_{i,j}-1}\alpha_{l} = \beta \\
        0 & {\rm otherwise}
    \end{cases} \qquad z_{i,j} > 1 \notag \\
    &\mathcal{T}_{(i,j)}(x_{i,j})_{\beta} = \delta_{\beta,0}, \qquad z_{i,j} = 1,
\end{align}
which will guarantee the tree tensor network contracts to $f(x)$, with the virtual indices all of dimension $\chi = d + 1$.
\subsection*{Mutual information measurement}
In Fig. \ref{fig:3DFunctionCompression} we plot the ``quantum mutual information" between two binary digits $x_{A} = x_{i,j}$ and $x_{B} = x_{i',j'}$ for the given function $f(\mathbf{x})$. To compute this quantity we treat the binary digits as spin degrees of freedom and the amplitudes of the function as analogous to the amplitudes of a quantum wavefunction $\psi (\mathbf{x}) \leftrightarrow f(\mathbf{x})$. We then form the two-body reduced density matrix $\rho_{A,B} = \rho(\left(x_{A}, x_{B}), (x_{A}, x_{B})'\right)$ by taking the partial trace over all bits excluding $x_{A}$ and $x_{B}$  of the full density matrix whose elements are specified as $\rho(\mathbf{x}, \mathbf{x}') = f(\mathbf{x})f^{*}(\mathbf{x}')/\sum_{\mathbf{y}}|f(\mathbf{y})|^{2}$. We have used the notation $\mathbf{x} = (x_{1,1}, x_{1,2}, \hdots x_{1,L}, x_{2,1}, x_{2,2}, \hdots x_{n,L})$ and $\mathbf{x}' = (x'_{1,1}, x'_{1,2}, \hdots x'_{1,L}, x'_{2,1}, x'_{2,2}, \hdots x'_{n,L})$.

If the function $f(\mathbf{x})$ is encoded as a tree tensor network, one can efficiently form the two-body reduced density matrix $\rho_{A,B}$ by forming a norm network from two copies of $f(\mathbf{x})$ and contracting over all physical indices other than $x_{A}$ and $x_{B}$. Such a contraction can always be  done in $\mathcal{O}(n\chi^{z+1} L)$ time (assuming $d=2$ for a binary digit) for any given two-body rdm and the error in the rdm is solely tied to the error in the TTN's representation of $f(\mathbf{x})$.

If the function is not in TTN format but is queriable, one can form an approximate reduced density matrix by performing the partial trace from sampling of the function over $N$ randomly selected grid points. For the example plotted in Fig. \ref{fig:3DFunctionCompression} we do this and find good convergence and $10^{4}$ grid points is sufficient to get an accurate approximation of the two-body reduced density matrix for a given pair of binary digits.

From this approximate matrix we can form the one-body reduced density matrices \newline $\rho_{A} = {\rm Tr}_{B}\rho_{A,B}$ and $\rho_{B} = {\rm Tr}_{A}\rho_{A, B}$, allowing the mutual information $M(x_{A}, x_{B})$ between the two bits to be calculated via
\begin{align}
M(x_A, x_B) &=  S(x_{A}) + S(x_{B}) - S(x_{A}\cup x_{B}) \notag \\
     &=  -{\rm Tr}\left(\rho_{A}\ln(\rho_{A})\right) - {\rm Tr}\left(\rho_{B}\ln(\rho_{B})\right) + {\rm Tr}\left(\rho_{A,B}\ln(\rho_{A,B})\right), 
     \label{eq:SMMutualInfo}
\end{align}
where $S(\cdot)$ denotes the standard Von-Neumann entropy.

\subsection*{Further TCI data}
\begin{figure}[t!]
    \centering
    \includegraphics[width = \columnwidth]{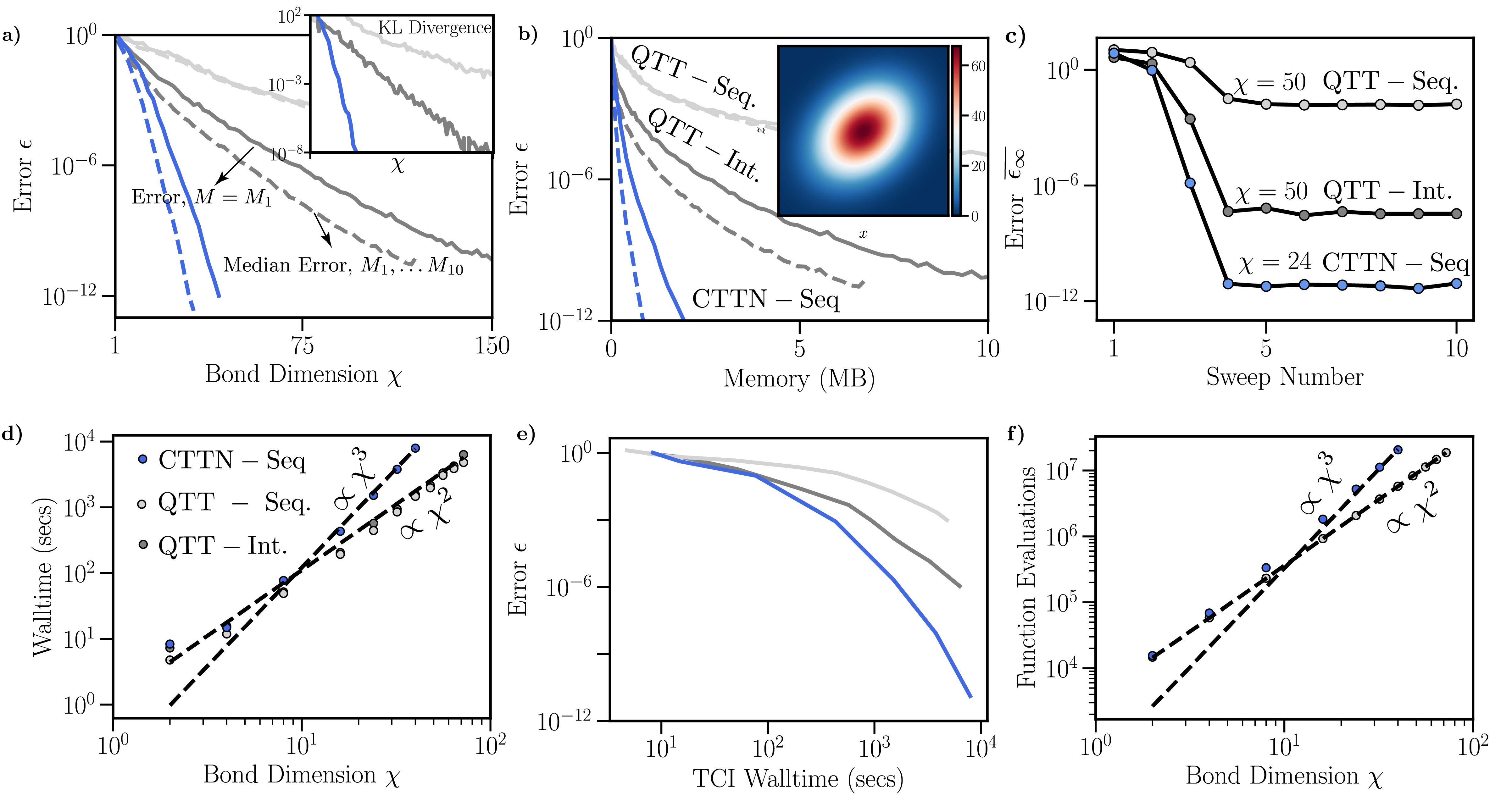}
    \caption{Comparison of the effectiveness of different tree tensor networks (diagrams of the different networks can be found in Fig. \ref{fig:TCI3D}) with $L = 16$ bits per dimension for learning the multinormal probability density function $f(\mathbf{r}) \propto \exp( - ((\mathbf{r} - \mathbf{\mu})^{T} M^{-1} (\mathbf{r} - \mathbf{\mu}))$ via the tensor cross interpolation algorithm. Here $M$ is an $n \times n$ covariance matrix and $\mathbf{\mu} = (\mu_{1}, \mu_{2}, ..., \mu_{n})$ is the mean vector. We consider $n =3$ with $\mathbf{r} = (x,y,z) \in [0,10)^{3}$ and $\mathbf{\mu} = (5,5,5)$. Results are obtained from drawing $N = 10$ instances of $M$ ($M_{1}, M_{2}, \hdots M_{10}$) from the LKJ distribution \cite{Lewandoski2009} with shape parameter $\eta = 1$, a distribution uniform in the space of all covariance matrices.
    \textbf{a-b)} Error $\epsilon$, calculated via Eq. (\ref{Eq:Errors}) after $10$ sweeps of the TCI algorithm, versus bond dimension and memory cost for the tensor networks. 
    The dashed lines shows the mode of the error over the $10$ realizations of $M$ while the solid line shows the specific error for $M = M_{1}.$ Inset in \textbf{a)} shows the Kullback-Leibler divergence (calculated from the function samples) as a function of bond dimension evaluated from the function samples for $M = M_{1}$. Inset in \textbf{b)} shows a heatmap of the function for $y = 5$.
    \textbf{c)} Average value for the infinity norm $\epsilon_{\infty}$  --- see Eq. (\ref{Eq:Errors}) --- over a given sweep of the TCI algorithm for $M = M_{1}$ and the three tensor networks at the specified bond dimensions. \textbf{d-f)} Walltime versus bond dimension, Error $\epsilon$ vs Walltime and number of evaluations of the function $f$ versus bond dimension respectively. Dotted lines show polynomial fits with bond dimension with the annotated scaling.
    }
    \label{fig:SFTCI3D}
\end{figure}

In Fig. \ref{fig:SFTCI3D} we present further data comparing the effectiveness of the tensor cross interpolation (TCI) algorithm for learning the trivariate probability density function \newline $f(\mathbf{r}) \propto \exp( - (\mathbf{r} - \mathbf{\mu})^{T} M^{-1} (\mathbf{r} - \mathbf{\mu}))$ where $\mathbf{r} = (x,y,z)$, $\mathbf{\mu} = (\mu_{x}, \mu_{y}, \mu_{z})$ is the mean vector, and $M$ is a covariance matrix. Here we sample $M$ from the Lewandowski-Kurowicka-Joe (LKJ) \cite{Lewandoski2009} distribution with shape parameter $\eta = 1$, which is equivalent to sampling uniformly from the space of all covariance matrices. As in the main text, we find the comb tree systematically outperforms the tensor trains for a given covariance matrix $M$. This is most transparent in Fig. \ref{fig:SFTCI3D}c), where the TCI algorithm converges to a value for the infinity norm $\epsilon_{\infty}$ that is orders of magnitude lower for the comb tree than for the tensor trains - despite the train Ansatzes utilizing a higher bond dimension and memory.
Due to the lower value of the shape parameter $\eta$ we find the bond dimensions required to reach a certain error, however, vary much much more strongly with different covariance matrices than in Fig. \ref{fig:TCI3D} where we set $\eta = 50$.  This is because, for $\eta =1$, matrices are drawn with significant inter-dimensional correlations. These are in general more difficult to represent with a tensor network Ansatz due to the presence of significant inter-dimensional correlations between binary digits.
\par We also show walltime data of 10 sweeps of our TCI implementation versus bond dimension as well as the number of function evaluations versus bond dimension. We find the comb tree tensor network is superior in this regard too (achieving a given error in less walltime) than the tensor trains. Interestingly we find the scaling of walltime with bond dimensions follows $\chi^{3}$ and $\chi^{2}$ for the comb tree and tensor trains respectively. This implies that querying the function is the dominant cost in our TCI implementation for this problem and range of bond dimensions. The interpolative decomposition scales as $\chi^{4}$ and $\chi^{3}$ for the comb tree and tensor trains respectively and thus its scaling should dominate at high bond dimensions --- but clearly not in the range of bond dimensions considered here --- where the function evaluations are dominating. We emphasize that such an observation might be implementation dependent. More efficient caching of the function queries might bring the prefactor down significantly and expose the asymptotic ID scaling sooner.

\end{appendix}

\bibliography{main}
\nolinenumbers
\end{document}